\documentclass[11pt]{article}
\usepackage{graphicx}
\usepackage{amssymb}
\usepackage{amscd}
\usepackage{mathrsfs}
\usepackage{longtable,lscape}
\usepackage{amsthm}
\usepackage{amsfonts}
\usepackage{amsmath}
\usepackage{bbm}
\usepackage{float}
\usepackage{url}
\usepackage{csquotes}
\usepackage{wrapfig}

\usepackage{caption}
\usepackage{lineno}
\usepackage[title]{appendix}
\begin{document}
\title{\bf Violation of CHSH inequality and marginal laws in mixed sequential measurements with order effects}
\author{Massimiliano Sassoli de Bianchi\vspace{0.5 cm} \\ 
\normalsize\itshape
Center Leo Apostel for Interdisciplinary Studies, \\ \itshape Brussels Free University, 1050 Brussels, Belgium\vspace{0.2 cm} \\ 
\itshape
Laboratorio di Autoricerca di Base, \\ \itshape c/o Area 302, 6917 Barbengo, Switzerland\vspace{0.2 cm} \\
\normalsize
E-Mails: \url{msassoli@vub.ac.be}, \ \url{autoricerca@gmail.com}
}
\date{}
\maketitle
\begin{abstract} 
\noindent We model a typical Bell-test experimental situation by considering that Alice and Bob perform incompatible measurements in a sequential way, with mixed orders of execution. After emphasizing that order effects will generally produce a violation of the marginal laws, we derive an upper limit for the observed correlations. More precisely, when Alice's and Bob's measurements are compatible, the marginal laws are obeyed and Tsirelson's bound limits the quantum correlations in the Bell-CHSH inequality to $2\sqrt{2}$. On the other hand, when Alice and Bob perform incompatible mixed sequential measurements, the marginal laws are typically violated and the upper limit for the correlations is pushed up to $2\sqrt{3}$. Considering that significant violations of the marginal laws (also called no-signaling conditions) have been observed in the data of numerous Bell-test experiments, the present analysis provides a possible mechanism for their appearance, when the protocols are such that Alice's and Bob's measurements can be assumed to be performed in a mixed sequential way. We however emphasize that this does not imply that a communication with superluminal effective speed would be possible.
\end{abstract} 

\section{Introduction}
\label{Introduction}

In a typical two-channel Bell-test experiment, a bipartite entity in a pre-measurement state $|\psi\rangle\in {\cal H} \simeq {\cal H}_1\otimes{\cal H}_2$ is submitted to four different joint measurements, described by the tensor product observables: 
\begin{equation}
A\otimes B,\quad A'\otimes B,\quad A\otimes B',\quad A'\otimes B', 
\label{product-measurements}
\end{equation}
where $A$ and $A'$ are self-adjoint operators acting on the state space ${\cal H}_1$ of the first sub-entity (say, that measured by Alice's apparatuses), whereas $B$ and $B'$ are self-adjoint operators acting on the state space ${\cal H}_2$ of the second sub-entity (say, that measured by Bob's apparatuses). The paradigmatic situation is when the entity in question is formed by two spin-${1\over 2}$ sub-entities, hence ${\cal H}_1 = {\cal H}_2=\mathbb{C}^2$, and the four one-entity operators $A$, $A'$, $B$ and $B'$ are spin observables corresponding to different spatial directions (i.e., different orientations of the Stern-Gerlach apparatuses) and the pre-measurement spin state is for instance the rotationally invariant singlet state: 
\begin{equation}
|\psi\rangle ={1\over\sqrt{2}}(|+\rangle\otimes|-\rangle - |-\rangle\otimes|+\rangle).
\label{singlet}
\end{equation}
The presence of entanglement can be tested via the violation of the Bell-CHSH inequality \cite{CHSH1969}. For this, one considers that the four one-entity operators $A$, $A'$, $B$, $B'$ are normalized in such a way that their eigenvalues are $\pm 1$ (this is the case for Pauli's matrices). Hence, one can write the spectral decompositions (with obvious notation): 
\begin{eqnarray}
&&A=P_{A_+}-P_{A_-}=\mathbb{I}-2 P_{A_-},\quad A'=P_{A'_+}-P_{A'_-}=\mathbb{I}-2 P_{A'_-}\nonumber\\
&&B=P_{B_+}-P_{B_-}=\mathbb{I}-2 P_{B_-},\quad B'=P_{B'_+}-P_{B'_-}=\mathbb{I}-2 P_{B'_-}
\label{ABA'B'}
\end{eqnarray}
Then, the Bell-CHSH inequality affirms that if only classical correlations are observable by the coincidence  measurements, then the absolute value of the average $\langle \psi |C|\psi\rangle$, of the observable:
\begin{equation}
C= A\otimes B -A\otimes B' + A'\otimes B' + A'\otimes B,
\label{standard-C}
\end{equation}
or of similar observables obtained by interchanging the roles of $A$ and $A'$ and/or $B$ and $B'$, will be less or equal than $2$, i.e., $|\langle \psi |C|\psi\rangle|\leq 2$. On the other hand, if the inequality is violated, then the observed correlations are said to be non-classical, or quantum. To use Aerts' terminology \cite{Aerts1990}, correlations violating the Bell-CHSH inequality are not `of the first kind', i.e, are not correlations of the kind that are already present before the measurements are executed (like in the famous example of Bertlmann's socks \cite{Bell1980}), but `correlations of the second kind', which are literally created by and during the measurements. 

Now, because of the tensor product structure, we can observe that: 
\begin{equation}
A\otimes B = (A\otimes \mathbb{I})(\mathbb{I}\otimes B)=(\mathbb{I}\otimes B)(A\otimes \mathbb{I}).
\label{AB-commutation}
\end{equation}
This means that $A\otimes B$ is the product of two compatible observables, and the same is obviously true also for the other observables forming the operator $C$. Consequently, a measurement of $A\otimes B$ can be equivalently interpreted as a joint (pure coincidence) measurement, or as a sequential measurement where, say, $A\otimes \mathbb{I}$ is executed first and is then followed by the execution of $\mathbb{I}\otimes B$, or the other way around. Indeed, in view of the commutation relation (\ref{AB-commutation}), there cannot be order effects and the fact that Alice's and Bob's measurements are performed in sequence, in whatever order, or in perfect simultaneity, the obtained statistics of outcomes will not be affected by that (see \cite{Khrennikov2011} for a critical discussion of the quantum postulate of simultaneous measurement of compatible observables). 

Another consequence of the tensor product structure of the four observables (\ref{product-measurements}), is that the outcome probabilities of their measurements will automatically obey the `marginal laws' (also called `marginal selectivity' or `no-signaling conditions'), stating that the probabilities obtained by Alice do not depend on the measurements performed by Bob, and vice versa. More precisely, denoting ${\cal P}_\psi(A_i,B_j)=\langle \psi |P_{A_i}\otimes P_{B_j}|\psi\rangle$ the probability of the outcome $(A_i,B_j)$ in measurement $A\otimes B$, for the pre-measurement state $|\psi\rangle$, and denoting ${\cal P}_\psi(A_i,B'_j)=\langle \psi |P_{A_i}\otimes P_{B'_j}|\psi\rangle$ the probability of the outcome $(A_i,B'_j)$ in measurement $A\otimes B'$, $i,j=\pm$, we have:
\begin{eqnarray}
&&\sum_{j=\pm}{\cal P}_\psi(A_i,B_j)=\langle \psi |P_{A_i}\otimes (P_{B_+}+P_{B_-})|\psi\rangle = \langle \psi |P_{A_i}\otimes \mathbb{I}|\psi\rangle, \label{left}\\
&&\sum_{j=\pm}{\cal P}_\psi(A_i,B'_j)=\langle \psi |P_{A_i}\otimes (P_{B'_+}+P_{B'_-})|\psi\rangle = \langle \psi |P_{A_i}\otimes \mathbb{I}|\psi\rangle,\label{right}
\end{eqnarray}
hence (\ref{left}) is equal to (\ref{right}). Reasoning in the same way with the other joint probabilities, we thus obtain the four equalities: 
\begin{eqnarray}
&&\sum_{j=\pm}{\cal P}_\psi(A_i,B_j)=\sum_{j=\pm}{\cal P}_\psi(A_i,B'_j),\quad \sum_{j=\pm}{\cal P}_\psi(A'_i,B_j)=\sum_{j=\pm}{\cal P}_\psi(A'_i,B'_j),\nonumber\\
&&\sum_{i=\pm}{\cal P}_\psi(A_i,B_j)=\sum_{i=\pm}{\cal P}_\psi(A'_i,B_j),\quad \sum_{i=\pm}{\cal P}_\psi(A_i,B'_j)=\sum_{i=\pm}{\cal P}_\psi(A'_i,B'_j),
\label{marginal}
\end{eqnarray}
which constitute the marginal laws, or no-signaling conditions, automatically obeyed by the four product observables (\ref{product-measurements}). However, tests conducted to date in different laboratories highlight significant violations of these conditions; see for instance \cite{AdenierKhrennikov2007,DeRaedt2012,DeRaedt2013,AdenierKhrennikov2016,Bednorz2017,Kupczynski2017}. The origins of these violations is still the object of debate: they might just be experimental artefacts, or instead they might reveal that the observed processes of creation of correlations cannot be properly modeled by assuming the measurements to be of the product kind. Note however that despite their name, a violation of the no-signaling conditions does not necessarily imply that relativity would be violated, as we will explain in the final section of the present article, the purpose of which is twofold. To begin with, in Section~\ref{mixed}, we show that incompatible mixed sequential measurements can possibly explain the observed violations of the no-signaling conditions. Then, in Section~\ref{generalized}, we show that for such kind of measurements Tsirelson's bound does not hold anymore and has to be replaced by a different upper limit, which is however not the maximum algebraic limit. Finally, in Section~\ref{Conclusion}, we discuss our findings and add some final remarks.

\section{Mixed sequential measurements}
\label{mixed}

We consider  the situation where Alice's (respectively, Bob) sub-measurements are not of the product form $A\otimes \mathbb{I}$ and $A'\otimes \mathbb{I}$ (respectively, $\mathbb{I}\otimes B$ and $\mathbb{I}\otimes B'$). This means that Alice and Bob measurements are not assumed to be necessarily compatible, so that the order of their execution will have an effect on the statistics of outcomes. In other words, we drop the assumption that Alice's and Bob's actions, performed at arbitrarily large spatial distances from each other, would only have local effects, and assume instead that they would be able to operate at the level of the overall composite entity, because of its non-spatial nature (hence without their overall actions to be understood as ``spooky'' -- superluminal -- influences at a distance, propagating through space). 

Of course, also when measurements are of the product form, Alice's and Bob's actions are \emph{strictu sensu} affecting both sub-entities,  because of the presence of entanglement described at the level of the state (\ref{singlet}). Here, however, we assume that the change of state produced by Alice's and Bob's individual measurements is such that it cannot  be modeled by using the same tensor product structure for all four joint measurements (\ref{product-measurements}), and of course that Alice's and Bob's outcomes are actualized in a genuine (uniformly mixed) sequential way.\footnote{Another assumption would however be possible: that Alice's and Bob's measurements would be part of bigger joint measurements that cannot be decomposed into sequential sub-measurements \cite{AertsSozzo2014a,AertsSozzo2014b,AetAL2019}.}  

So, not only we have $[A,A']\neq 0$ and $[B,B']\neq 0$, as is usually assumed in Bell-test experiments, but also that $[A,B]\neq 0$, $[A,B']\neq 0$, $[A',B]\neq 0$ and $[A',B']\neq 0$, i.e., Alice's measurements are not necessarily compatible with Bob's measurements. The spectral decompositions (\ref{ABA'B'}) is still valid, with the difference that the projections belonging to Alice's and Bob's spectral families will not anymore mutually commute. 

We start by considering the probability ${\cal P}_\psi(A_i\!\to\! B_j)$ of obtaining outcome $(A_i,B_j)$, $i,j=\pm$, when performing first measurement $A$, then immediately after measurement $B$. It is given by: 
\begin{equation}
{\cal P}_\psi(A_i\!\to\! B_j)=\langle\psi|P_{A_i}P_{B_j}P_{A_i}|\psi\rangle.
\label{seq-prob-AB1}
\end{equation}
Indeed, according to the Born rule, the probability of obtaining $A_i$, following Alice's measurement $A$, is ${\cal P}_\psi(A_i)=\langle\psi|P_{A_i}|\psi\rangle$. Also, according to the projection postulate, the measurement will then produce the state transition:
\begin{equation}
|\psi\rangle \to |\psi_{A_i}\rangle={P_{A_i}|\psi\rangle\over \langle\psi|P_{A_i}|\psi\rangle^{1\over 2}}.
\label{state-transition}
\end{equation}
So, following the $A$-measurement (assumed here to be non-destructive), the state will be $|\psi_{A_i}\rangle$, hence, using once more the Born rule, the probability of obtaining $B_j$, when measurement $B$ is performed immediately after measurement $A$, conditional to the fact that the outcome $A_i$ was obtained, is: 
\begin{equation}
{\cal P}_\psi(B_j|A_i)=\langle\psi_{A_i}| P_{B_j}|\psi_{A_i}\rangle={\langle\psi|P_{A_i}P_{B_j}P_{A_i}|\psi\rangle\over \langle\psi|P_{A_i}|\psi\rangle}.
\label{conditional-probability}
\end{equation}
From (\ref{conditional-probability}), the probabilistic interpretation of (\ref{seq-prob-AB1}) follows, as is clear that we have: 
\begin{equation}
{\cal P}_\psi(A_i\!\to\! B_j)={\cal P}_\psi(B_j|A_i){\cal P}_\psi(A_i).
\end{equation}
Reasoning in the same way for the sequence where measurement $B$ is performed before measurement $A$, we also find: 
\begin{equation}
{\cal P}_\psi(B_j\!\to\! A_i)=\langle\psi|P_{B_j}P_{A_i}P_{B_j}|\psi\rangle,
\label{seq-prob-AB2}
\end{equation}
and assuming that the exact order of the sequence, at each run of the experiment, cannot be controlled by Alice and Bob, we can define the joint probability ${\cal P}_\psi(A_i,B_j)$, of obtaining outcome $(A_i,B_j)$, by considering the uniform average: 
\begin{eqnarray}
{\cal P}_\psi(A_i,B_j) &=& {1\over 2} [{\cal P}_\psi(A_i\!\to\! B_j)+ {\cal P}_\psi(B_j \!\to\! A_i)]\nonumber\\
&=& {1\over 2}\langle\psi|(P_{A_i}P_{B_j}P_{A_i}+ P_{B_j}P_{A_i}P_{B_j})|\psi\rangle.
\label{joint-average-prob}
\end{eqnarray}
Summing over $j$, we thus find: 
\begin{equation}
\sum_{j=\pm}{\cal P}_\psi(A_i,B_j) = {1\over 2} {\cal P}_\psi(A_i) + {1\over 2} \sum_{j=\pm} {\cal P}_\psi(B_j\!\to\! A_i),
\label{marginal1}
\end{equation}
so that the difference of the marginal probabilities is: 
\begin{equation}
\sum_{j=\pm}{\cal P}_\psi(A_i,B_j) -\sum_{j=\pm}{\cal P}_\psi(A_i,B'_j)={1\over 2}\sum_{j=\pm}\langle\psi|P_{B_j}P_{A_i}P_{B_j}|\psi\rangle - {1\over 2}\sum_{j=\pm}\langle\psi|P_{B'_j}P_{A_i}P_{B'_j}|\psi\rangle
\label{marginal-deviation}
\end{equation}
and similar expressions can be derived for the other combinations of Alice's and Bob's measurements. It is clear from the above that there can be a violation of the marginal laws (i.e., the r.h.s. of (\ref{marginal-deviation}) will be in general different from zero) if Alice's measurements are not compatible with Bob's measurements.

\section{An upper bound for the CHSH inequality}
\label{generalized}

Having observed that uniformly mixed sequential measurements, if incompatible, allows for a violation of the marginal laws (the no-signaling conditions), we want now to derive a bound corresponding to the maximal violation of the Bell-CHSH inequality in this case. Such bound will play the same role as Tiserlon's bound, which only applies in the situation where Alice's and Bob's measurements are compatible, which is always the case if their joint measurements are described by the tensor products observables (\ref{product-measurements}). Here we relax this requirement, considering the hypothesis that what happens in the laboratory can be conveniently described in terms of a uniform mixture of sequential incompatible measurements performed in different orders. Our starting point is the CHSH quantity:
\begin{equation}
{\rm CHSH}_\psi = E_\psi(A,B)-E_\psi(A,B') + E_\psi(A',B') + E_\psi(A',B),
\label{B-CHSH}
\end{equation}
where we have defined the four averages (also called `correlation functions'): 
\begin{eqnarray}
&&E_\psi(A,B)=\sum_{i,j=\pm} ij\, {\cal P}_\psi(A_i,B_j),\quad E_\psi(A,B')= \sum_{i,j=\pm} ij\, {\cal P}_\psi(A_i,B'_j),\nonumber\\
&&E_\psi(A',B)=\sum_{i,j=\pm} ij\, {\cal P}_\psi(A'_i,B_j),\quad E_\psi(A',B')= \sum_{i,j=\pm} ij\, {\cal P}_\psi(A'_i,B'_j).
\label{averages}
\end{eqnarray}
We explicitly calculate only $E_\psi(A,B)$, the calculation for the other averages in (\ref{B-CHSH}) being similar. Since we are here considering mixed sequential measurements, 
we have to use the probabilities (\ref{joint-average-prob}). This gives: 
\begin{eqnarray}
E_\psi(A,B)&=& {\cal P}_\psi(A_+,B_+) - {\cal P}_\psi(A_+,B_-) + {\cal P}_\psi(A_-,B_-) - {\cal P}_\psi(A_-,B_+)\\ \nonumber
&=& {1\over 2}\langle\psi|(P_{A_+}P_{B_+}P_{A_+}+ P_{B_+}P_{A_+}P_{B_+})|\psi\rangle - {1\over 2}\langle\psi|(P_{A_+}P_{B_-}P_{A_+}+ P_{B_-}P_{A_+}P_{B_-})|\psi\rangle\\ \nonumber
&+& {1\over 2}\langle\psi|(P_{A_-}P_{B_-}P_{A_-}+ P_{B_-}P_{A_-}P_{B_-})|\psi\rangle- {1\over 2}\langle\psi|(P_{A_-}P_{B_+}P_{A_-}+ P_{B_+}P_{A_-}P_{B_+})|\psi\rangle
\\ \nonumber
&=&{1\over 2}\langle\psi|(P_{A_+}BP_{A_+}+ P_{B_+}AP_{B_+}-P_{B_-}AP_{B_-}-P_{A_-}BP_{A_-})|\psi\rangle,
\label{average-E}
\end{eqnarray}
where for the last equality we have used (\ref{ABA'B'}). Further replacing $A$ and $B$ by $\mathbb{I}-2P_{A_-}$ and $\mathbb{I}-2P_{B_-}$, respectively, we find for the above term in brackets:
\begin{eqnarray}
(\,\cdots)&=& P_{A_+}(\mathbb{I}-2P_{B_-})P_{A_+}+ P_{B_+}(\mathbb{I}-2P_{A_-})P_{B_+}-P_{B_-}(\mathbb{I}-2P_{A_-})P_{B_-}-P_{A_-}(\mathbb{I}-2P_{B_-})P_{A_-}\nonumber\\
&=&P_{A_+}- 2P_{A_+}P_{B_-}P_{A_+}+ P_{B_+}-2P_{B_+}P_{A_-}P_{B_+}-P_{B_-} +2P_{B_-}P_{A_-}P_{B_-}-P_{A_-} +2P_{A_-}P_{B_-}P_{A_-}\nonumber\\
&=&(A + B) +2P_{A_-}P_{B_-}P_{A_-} - 2P_{A_+}P_{B_-}P_{A_+}+2P_{B_-}P_{A_-}P_{B_-} -2P_{B_+}P_{A_-}P_{B_+}.
\label{dots}
\end{eqnarray}
We can also observe that:
\begin{eqnarray}
&&P_{A_-}P_{B_-}P_{A_-} - P_{A_+}P_{B_-}P_{A_+}=(\mathbb{I}-P_{A_+})P_{B_-}(\mathbb{I}-P_{A_+}) - P_{A_+}P_{B_-}P_{A_+}\nonumber\\
&&=P_{B_-} -P_{A_+}P_{B_-}-P_{B_-}P_{A_+} =(\mathbb{I}-P_{A_+})P_{B_-} -P_{B_-}P_{A_+}=P_{A_-}P_{B_-}-P_{B_-}P_{A_+}.
\label{aba}
\end{eqnarray}
Similarly, we have: 
\begin{eqnarray}
&&P_{B_-}P_{A_-}P_{B_-} -P_{B_+}P_{A_-}P_{B_+}=(\mathbb{I}-P_{B_+})P_{A_-}(\mathbb{I}-P_{B_+}) - P_{B_+}P_{A_-}P_{B_+}\nonumber\\
&&=P_{A_-} -P_{B_+}P_{A_-}-P_{A_-}P_{B_+} =(\mathbb{I}-P_{B_+})P_{A_-} -P_{A_-}P_{B_+} =P_{B_-}P_{A_-}-P_{A_-}P_{B_+}.
\label{bab}
\end{eqnarray}
Inserting (\ref{aba}) and (\ref{bab}) into (\ref{dots}), gives: 
\begin{eqnarray}
(\,\cdots)&=&A + B +2P_{A_-}P_{B_-}-2P_{B_-}P_{A_+} + 2P_{B_-}P_{A_-}-2P_{A_-}P_{B_+} \nonumber\\
&=&A + B -2P_{B_-}A -2P_{A_-}B = (\mathbb{I} -2P_{B_-})A + (\mathbb{I} -2P_{A_-})B\nonumber\\
&=&BA + AB.
\end{eqnarray}
We thus find:
\begin{equation}
E_\psi(A,B)=\langle\psi|\widehat{AB}|\psi\rangle, 
\label{E-bis}
\end{equation}
where we have defined the symmetrized self-adjoint operator 
\begin{equation}
\widehat{AB}\equiv {1\over 2}[AB + (AB)^\dagger] = {1\over 2}(AB + BA),
\label{symmetrized}
\end{equation}
i.e., we find that the correlation function relative to the uniform mixing of the $A$ and $B$ measurements, performed in a sequential way, is given by the quantum average of the operator (\ref{symmetrized}), which of course reduces to the standard product $A\otimes B$ in the special case where Alice and Bob observables commute and are of the form $A\otimes \mathbb{I}$ and $\mathbb{I}\otimes B$, respectively. Similar expressions can be obtained in the same way for the other mixed sequential measurements, so that (\ref{B-CHSH}) can be written as: 
\begin{eqnarray}
&&{\rm CHSH}_\psi = \langle\psi|\widehat{C}|\psi\rangle, \quad \widehat{C}\equiv {1\over 2}(C + C^\dagger),\quad \widehat{C}^\dagger = \widehat{C},\\ \nonumber
&&C= AB-AB' + A'B' + A'B,\quad \widehat{C}= \widehat{AB}-\widehat{AB'} + \widehat{A'B'} +\widehat{A'B}.
\label{B-CHSH-2}
\end{eqnarray}

To obtain a bound on ${\rm CHSH}_\psi$, defined as per above, it is instructive to first study the operator $C$. Following Khalfin and Tsirelson's algebraic method \cite{KhalfinTsirelson1985}, one proceeds by analyzing the different terms appearing when taking the square of $C$. One finds: 
\begin{equation}
C^2 = C_1 + C_2 + \Delta_1 + \Delta_2 + \Delta_3 + \Delta_4, 
\end{equation}
where we have defined: 
\begin{eqnarray}
C_1 &=& (AB)^2 +(AB')^2+(A'B')^2+ (A'B)^2, \nonumber \\
C_2 &=& ABA'B' - AB'A'B + A'B'AB -A'BAB',\nonumber \\
\Delta_1 &=& ABA'B -AB'A'B'=A[B,A']B -A[B',A']B', \nonumber \\
\Delta_2 &=& A'B'A'B - AB'AB= A'[B',A']B - A[B',A]B,\nonumber\\
\Delta_3 &=& A'BAB -A'B'AB' = A'[B,A]B -A'[B',A]B',\nonumber\\
\Delta_4 &=& A'BA'B'-ABAB' = A'[B,A']B'-A[B,A] B'.
\end{eqnarray}
We can observe that if Alice's and Bob's measurements are compatible, then the above commutators are zero and $\Delta_1=\Delta_2=\Delta_3=\Delta_4=0$. And since $A^2= B^2= {A'}^2 + {B'}^2 = \mathbb{I}$ (all these operators have $\pm 1$ eigenvalues), we also have $C_1 =4\mathbb{I}$. If Alice's observables $A$ and $A'$ are in addition also mutually compatible, and same for Bob's observables $B$ and $B'$, then $C_2=0$, so that in this case $C^2 = 4\mathbb{I}$, from which the classical bound of the CHSH inequality can be deduced. Let us do all the steps. By definition of the norm, we have: 
\begin{eqnarray}
\| C\|^2 = \sup_{\|\phi\|=1} \|C\phi\|^2 = \sup_{\|\phi\|=1}\langle C\phi |C\phi\rangle =\sup_{\|\phi\|=1}\langle \phi |C^\dagger C|\phi\rangle \leq \| C^\dagger C\|.
\label{norm}
\end{eqnarray}
If $C^\dagger =C$, which is the case here under our assumptions, one finds that $\|C\|^2\leq \|C^2\|$. But since we also have $\|C^2\|\leq \|C\|^2$, we can conclude that $\|C^2\|= \|C\|^2$. Finally, since $\langle \psi |C|\psi\rangle^2\leq \| C\|^2= \| C^2\|=4$, we obtain that $-2\leq \langle \psi |C|\psi\rangle \leq 2$, which is the CHSH inequality. In the quantum case, when the observables are of the tensor product form, Alice's and Bob's measurements are still compatible, but this time $[A,A']\neq 0$ and $[B,B']\neq 0$, hence $C_2 \neq 0$. The best one can then do is to observe that $\|C_2\| \leq 4 \| A\| \| A'\| \| B\| \| B' \| = 4$, so that $\|C^2\| =\| 4\mathbb{I} + C_2\| \leq 8$, hence this time the bound is: $-\sqrt{8}\leq \langle \psi |C|\psi\rangle \leq \sqrt{8}$, which is Tsirelson's bound for the quantum correlations. 

The situation we want now to analyze is such that all observables are in principle incompatible, but measurements are performed in a uniformly mixed way. Instead of $C$, we thus have to use the more general operator: $\widehat{C}= \widehat{AB}-\widehat{AB'} + \widehat{A'B'} +\widehat{A'B}$. We have:
\begin{eqnarray}
({\widehat C})^2&=& G_1 + G_2 + D_1 + D_2 +D_3 + D_4, \nonumber \\
G_1 &=& (\widehat{AB})^2 +(\widehat{AB'})^2+(\widehat{A'B'})^2+ (\widehat{A'B})^2,\nonumber \\
G_2 &=& \widehat{AB}\widehat{A'B'} - \widehat{AB'}\widehat{A'B} + \widehat{A'B'}\widehat{AB} -\widehat{A'B}\widehat{AB'},\nonumber \\
D_1 &=& \widehat{AB}\widehat{A'B} -\widehat{AB'}\widehat{A'B'},\nonumber \\
D_2 &=& \widehat{A'B'}\widehat{A'B} - \widehat{AB'}\widehat{AB},\nonumber\\
D_3 &=& \widehat{A'B}\widehat{AB} -\widehat{A'B'}\widehat{AB'},\nonumber\\
D_4 &=& \widehat{A'B}\widehat{A'B'}-\widehat{AB}\widehat{AB'}.
\end{eqnarray}
We observe that $\|\widehat{AB}\| = {1\over 2}\|AB + BA \|\leq {1\over 2}(\|AB\| + \|BA \|)\leq 1$, so that $\|(\widehat{AB})^2\|\leq \|\widehat{AB}\|^2 \leq 1$, 
and $\| G_1\| \leq 4$. Similarly, we have $\| G_2\| \leq 4$. Expanding the $D_1$ term, we find: 
\begin{eqnarray}
D_1 &=& \widehat{AB}\widehat{A'B} -\widehat{AB'}\widehat{A'B'} \nonumber\\
&=&{1\over 4}[(AB+BA)(A'B + BA')-(AB'+B'A)(A'B'+B'A')]\nonumber\\
&=&{1\over 4}[(ABA'B +BABA' +AA' + BAA'B)-(AB'A'B' + AA' +B'AA'B'+ B'AB'A')]\nonumber\\
&=&{1\over 4}[(ABA'B +BABA' + BAA'B)-(AB'A'B' +B'AA'B'+ B'AB'A')].
\end{eqnarray}
This means that: $\| D_1 \|\leq {6\over 4}$. Similar calculations show that 
\begin{eqnarray}
&&D_2 = {1\over 4}[(A'B'A'B + A'B'BA' + B'A'BA')-(AB'AB + AB'BA + B'ABA)],\nonumber\\
&&D_3 = {1\over 4}[(A'BAB +BA'AB + BA'BA)-(A'B'AB' +B'A'AB'+B'A'B'A)],\nonumber\\
&&D_4 = {1\over 4}[(A'BA'B' + A'BB'A' + BA'B'A')-(ABAB' + ABB'A+BAB'A)].
\end{eqnarray}
Hence, we also have $\| D_2 \|,\| D_3 \|,\| D_4 \|\leq {6\over 4}$.
Combining the above contributions, we thus conclude that $\|({\widehat C})^2\|\leq 8 + 4{6\over 4}=12$. Therefore, considering that ${\widehat C}$ is self-adjoint, we also have (reasoning as above) $\|{\widehat C}\|^2= \|({\widehat C})^2\|$, from which we deduce the bound: 
\begin{equation}
-2\sqrt{3} \leq \langle\psi|\widehat{C}|\psi\rangle \leq 2\sqrt{3}.
\label{theorem}
\end{equation}

The above constitutes the quantum upper limit for the correlations that can be generated by uniformly mixing incompatible sequential measurements. Note that $2\sqrt{3}\approx 3.46<4$, with $4$ being the algebraic maximum value attainable by the CHSH quantity (\ref{B-CHSH}). Hence, mixed sequential measurements do not fill the entire gap between 2 and 4, but are limited by the value $2\sqrt{3}$, which generalizes Tsirelson's bound of $2\sqrt{2}$, the latter being only valid when Alice's and Bob's measurements are all compatible and therefore the marginal laws are necessarily obeyed. 

Note that in the limit situation where $A',B'\to \mathbb{I}$, i.e, where the primed measurements are trivial measurements, then ${\widehat C}\to{1\over 2}(AB + BA) +\mathbb{I} +B-A$, and it is again a matter of some simple algebra to check that $\|({\widehat C})^2\|\leq 8$, i.e., Tsirelson's bound holds in this limit situation. 
Note also that in the limit where, say, $A'\to A$, i.e., where Alice always perform the same measurement, one can easily check that: $\|({\widehat C})^2\|\leq 4$, i.e., the classical bound holds. On the other hand, it remains an open question that of determining if the $2\sqrt{3}$ bound is tight for the general situation, or if it can be further optimized.

\section{Discussion}
\label{Conclusion}

We observed that a uniform mixing of sequential measurements performed by Alice and Bob corresponds to considering the four effective observables: 
\begin{equation}
\widehat{AB},\quad \widehat{A'B},\quad \widehat{AB'},\quad \widehat{A'B'}, 
\label{non-product-measurements}
\end{equation}
which replace the usual product observables (\ref{product-measurements}). In other words, by uniformly mixing two sequential measurement, say the $A$ and $B'$ measurements, everything happens as if Alice and Bob would jointly perform a single whole measurement, executed ``at once,'' corresponding to an observable of the form $\widehat{AB'}= {1\over 2}(AB' + B'A)$. It is the specific symmetrized form of this and the others effective observables that explains why the violation of the CHSH inequality can be bounded by the $2\sqrt{3}$ limit value. 

Clearly, for measurements that would be jointly performed by Alice and Bob, not in a consecutive way, described by generic non-product observables, the CHSH inequality can in principle be violated up to its maximum algebraic value. In fact, our analysis precisely tells us that if the violation is beyond $2\sqrt{3}$, and measurements can be conveniently modeled within the Hilbertian formalism, then they certainly are not describable as mixed sequential measurements. An example of this are the measurements described in Aerts' connected vessels of water model \cite{Aerts1982}, a macroscopic entity on which measurements can be defined that are able to maximally violate the Bell-CHSH inequality, i.e., with value 4, modelizable in Hilbert space by introducing general non-product observables \cite{AertsSozzo2014a,AertsSozzo2014b,AetAL2019}, not of the symmetrized form (\ref{non-product-measurements}). 

Note however that if we relax the hypothesis that measurements can be modeled within the standard formalism of quantum mechanics, i.e., by means of \emph{bona fide} self-adjoint operators, then we can also have situations of uniformly mixed sequential measurements that can violate the CHSH inequality beyond the $2\sqrt{3}$ limit, an example being the `generalized rigid rod model' analyzed in \cite{AetAL2018}. 

We can observe that if experiments are performed in such a way that Alice's and Bob's detection instruments are placed at different distances from the source of the bipartite entity, they will not click in coincidence but with a given time-delay, proportional to the difference in flight times. In this way, by considering a sufficiently large time-delay, one can make sure that the measurements will be genuinely sequential. By comparing the statistics of outcomes obtained by mixing the orders of sequential detection, with that obtained when Alice's and Bob's apparatuses are at exactly the same distance from the source (hence their outcomes can be actualized within a very narrow coincidence time interval), one can in principle check if these two descriptions -- mixed sequential and simultaneous -- are equivalent or not (see \cite{Zbindenetal2001} for an experiment where the detectors move, so that in a given reference frame they do not meet the measured entity at the same moment). In particular, it would be also interesting to see if there can be differences in the way the marginal laws are possibly violated in these two distinct experimental situations. 

As regards the violations of the or marginal laws, or no-signaling conditions, as we said they have been observed in several experiments performed to test
various Bell-type inequalities \cite{AdenierKhrennikov2007,DeRaedt2012,DeRaedt2013,AdenierKhrennikov2016,Bednorz2017,Kupczynski2017}, and their origin remains to be clarified. If in the end it will be shown that they are only experimental errors, they would of course not constitute a violation of the Einsteinian no-signaling principle. The other possibility would be that in real experiments some additional non-local processes could be at play that have not yet been identified, which considering the model we proposed in this article could be explained as order effects produced by mixed sequential measurements (see also \cite{AetAL2019}, where the violation of the marginal laws is explained instead in terms of joint measurements that are genuinely coincident). But even in this case, relativistic causality would not be necessarily at stake. Indeed, what is usually not taken into consideration in the standard analysis of the situation (see for instance \cite{Ballentine1987}) is that for superluminal signaling to be possible in practice, it is not sufficient to have the collapse of entangled states to be instantaneous and independent of the distance separating Alice's and Bob's locations, as this is not the only process involved in a communication. To define the effective time of the latter, one needs to consider how much time is needed, say for Alice, to initiate the emission of the required entangled state (in fact, of an entire statistical ensemble of them), by sending a signal to the source, then the additional time needed for the emitted entangled entity to reach Alice's and Bob's laboratories. When considering all these processes, Alice's effective messaging speed will probably slow down in such a way that signaling faster than light will be excluded. Last but not least, there is also the issue that in quantum measurements one cannot control the actualization in time of an outcome. Certainly, this would require a more specific analysis, which is however beyond  the scope of the present paper (see however the rather detailed related discussion presented in \cite{AetAL2019}). All we wanted to emphasize here is that a possible violation of the no-signaling conditions does not automatically lead to the possibility of a supraluminal communication: each experimental situation needs to be examined in a very careful way, so as to determine how much time actually elapses between the beginning and the end of a communication, and how these start and end times are to be defined at a fundamental level. 

A question we have not addressed in this paper is what kind of ``new'' physics could be behind the possibility for Alice's measurements to affect the marginal probabilities calculated by Bob in his measurements, and vice versa (assuming here that such possibility would not be a mere experimental artefact), i.e., the possibility that despite an arbitrarily large spatial separation of their apparatuses, Alice's and Bob's local (in space) measurements would remain mutually incompatible. We don't have a simple answer to this question. Let us consider however the following cognitive metaphor. Assume that you ask a group of persons if they believe Bill Clinton is honest and trustworthy, and immediately following such question you ask these same persons if they think Al Gore is honest and trustworthy (see for instance \cite{WangBusemeyer2013}). From all their answers, you can calculate the probabilities for the four possible outcomes. However, if you ask these two same questions in reverse order, the probabilities of the outcomes will be different, i.e., order effects will manifest. So, even though one question is only specific to the `Al Gore conceptual entity', and the other question to the `Clinton conceptual entity', in the mind of the respondents the process of answering the first question produces an instantaneous change in the context of the second one, hence affecting the way it will be subsequently processed. So, interpreting these two questions as the two measurements $A$ and $B$ that Alice and Bob perform on the two distinct conceptual sub-entities forming the ``Bill Clinton-Al Gore'' bipartite system, we have here a situation of incompatible measurements performed in a sequence, on a bipartite entity, which can mimic what might possibly happen (mutatis mutandis) with physical entities, when the no-signaling conditions are violated (see \cite{Aertsetal2018} for a speculative view where quantum entities are assumed to share with
human concepts a similar conceptual nature, which would explain among other things their non-local behavior). 

Based on our classical prejudices, we would tend to dismiss the existence of connections between measuring apparatuses when they are clearly spatially separated. However, with the discovery of entanglement, we certainly had to update these prejudices and accept that apparatuses that are spatially separated are not necessarily for this also experimentally separated. In fact, the observed violations of the Bell-CHSH inequality were precisely about showing that some kind of non-spatial connection must exist that is responsible for the observed correlations that measurements can create \cite{AertsSassoli2016}. The mechanism at the origin of these correlations is usually  described, in the quantum formalism, as something to be associated with the state of the measured bipartite system, and not as something to be included in the description of the observables. However, it is also known that the quantum formalism allows in principle for the entanglement resource to be totally or partially shifted from the state to the observables \cite{Harshman2011}, so one cannot exclude that the non-spatial phenomenon of entanglement could jointly manifest, in certain situations, not only in the states, but also in the accessible interactions, i.e., the measurements, which might be operated locally in space, but not for this would be local for what concerns their reach, i.e., their ability to act at the level of the overall bipartite entity.

This was recently suggested by Aerts and Sozzo in the entanglement scheme they proposed for modeling the dynamics of concepts and their combinations \cite{AertsSozzo2014a,AertsSozzo2014b} (see also the more recent analysis in \cite{AetAL2019}), showing that the data obtained in psychological measurements, systematically violating the marginal laws, require the introduction of entanglement not only at the level of the states, but also of the joint measurements. In other words, similarly to the effective observables (\ref{non-product-measurements}), which we introduced by assuming that Alice and Bob perform incompatible measurements, they also consider the necessity of introducing non-product (entangled) measurements, to properly describe the different joint actions that are performed by Alice and Bob. However, different from what we did in this paper, they consider experimental situations where Alice and Bob perform pure joint (simultaneous) measurements, operated ``at once'' by the individual minds of the  participants. For these experimental situations, which in a sense are more general than those considered here, the magnitude of the correlations in the Bell-CHSH inequality can only be limited by the maximum algebraic bound. 

To conclude, another question we have not addressed in this paper is how to possibly fundamentally define non-commuting observables replacing the usual product observables in specific experimental situations, when the no-signaling conditions are apparently violated. To answer this question would require a discussion of the mechanisms that could be at the origin of the deviations from the standard situation of compatibility, i.e., of commutability of Alice's and Bob's measurements, when separated by large spatial distances. Our much more modest intent in this paper was to show that the assumption of non-commutability, combined with that of a uniform mixing of sequential measurements (assuming that the experimental situation does actually involve genuine sequential measurements), allows for a modeling of probabilistic data exhibiting some violation of the no-signaling conditions, and also comes with an upper limit for the obtainable correlations. Our analysis can certainly be of interest in the quantum modeling of question order effects in psychological experiments \cite{BusemeyerBruza2012,asdb2015a}. As for knowing if it will be relevant also for the modeling of physics' experiments, in the study of quantum entanglement, this we will only know to the extent that more experiments will be conducted, and more reliable data will become available.

\section*{Acknowledgments} I thank with pleasure Sandro Sozzo for his useful comments in relation to the content of this article.

\end{document}